\documentclass[prb,amsmath,amssymb]{revtex4}

\usepackage{graphicx}% Include figure files
\usepackage{bm}% bold math
\usepackage{color}% bold math
\usepackage{rotating}
\usepackage{psfrag}
\usepackage{epsfig}

\begin{document}

%\wstoc{For Proceedings Contributors: Using World Scientific's\\ WS-PROCS9x6 Document Class in \LaTeX2e}{S.Aubry and L.Proville}

\title{{\large\sf Pressure Fronts in 1D Damped  Nonlinear Lattices }}

\author{S. AUBRY$^*$}
\affiliation{Laboratoire L\'eon Brillouin, CEA Saclay, 91191-Gif-sur-Yvette, France \\
$^*$E-mail: serge.aubry@cea.fr}

\author{L. PROVILLE}

\affiliation{SRMP, CEA Saclay, 91191-Gif-sur-Yvette, France}

\begin{abstract}
The propagation of pressure fronts
(impact solutions) in 1D chains of atoms coupled by  anharmonic potentials
between nearest neighbor and submitted to damping forces preserving uniform motion, is investigated.
Travelling  fronts between two regions at different uniform pressures are found numerically and well approximate analytically. It is proven that there are three analytical relations between the impact velocity, the compression, the front velocity and the energy dissipation  which only depend on the coupling potential and are \textit{independent}  of the damping.   Such travelling front solutions cannot exist without damping.
\end{abstract}


\maketitle
%\bodymatter

\section{The model}

This study was motivated for  understanding sonoluminescence
(see refs. in \cite{DA06})  observed for example when water saturated with rare gas
is submitted to an intense ultrasonic field.  Spherical bubbles
of rare gas expand and collapse periodically at supersonic or nearly supersonic speed  emitting  simultaneously a short and intense broadband  light pulse at impact on the bubble core
(sonoluminescence). We suggested more generally that  light emission is systematically generated at strong enough impacts  still below the range for generating plasmas \cite{DA06} but providing they become highly nonlinear.  This situation occurs  when the hard core of the atoms (which  prevent volume elements to become negative), is involved that is for supersonic or nearly supersonic impacts.
We suggest now a new remark  for amending our early theory.
Since condensed matter is made of  bonded charged particles, sharp accelerations of  these charges at strong enough impact  may be sufficient to generate  an intense  (Abrahams-Lorentz) electromagnetic (em) radiation
 visible as sonoluminescence and generating damping (note that a charged hard sphere model would produce a diverging radiation!).  We do not discuss here the physical validity of this suggestion  but   we only focus  on  some preliminary mathematical aspects of shocks  in  simple 1D model with damping.

FPU lattices are 1D chains of atoms coupled by anharmonic springs with Hamiltonian
\begin{equation}
\mathcal{H}= \sum_n  \left(\frac{1}{2} p_n^2 +\mathcal{V} (u_{n+1}-u_n)\right)
\label{FPU}
\end{equation}
where   $u_{n}(t)$ is the scalar coordinate of atom $n$,
$p_n=\dot{u}_n$ is the associated conjugate variable  and
 $\mathcal{V}(v)$ is the coupling  potential which depends on the  distance
$v_n=u_{n+1}-u_n$ between nearest neighbor atoms  $n+1$ and $n$.

We are interested in moving pressure fronts between two regions at different pressure obtained from initial conditions corresponding to an impact which are for example:
 $u_n(0)=nv$ and   $\dot{u}_n(0)= -V_P$  for $n>0$, $\dot{u}_n(0)= +V_P$ for $n<0$ and  $v_0(0)=0$.
 The chain at equilibrium is initially moving uniformly for positive $n$
and at the opposite velocity for negative $n$ ($V_P$ is the impact velocity).
Since by symmetry arguments, atom $0$ remains immobile  at all time ($u_0(t)=0$),
it is equivalent to consider that the positive part of the chain $n>0$) impacts a
fixed rigid wall at $u_0(0)=0$ or in the framework of the center of mass of the half chain that
 atom $0$ in the chain initially at rest is pushed by a piston  $u_0(t)= V_P t$. Studies were already devoted to pressure fronts in lattice models without damping  \cite{HL78,HFM81,Sto05}.
The same model with  damping preserving uniform motion
(as  in \cite{HKM95} for a Toda potential)  is described by  eqs.
\begin{eqnarray}
\ddot{u}_n &-& \gamma (\dot{u}_{n+1}+\dot{u}_{n-1}-2\dot{u}_n)-
\mathcal{V}^{\prime}(u_{n+1}-u_n)+\mathcal{V}^{\prime}(u_n-u_{n-1})=0
\qquad \mbox{or}\label{dyndamp1} \\
\ddot{v}_n &	&- \gamma \Delta. \dot{v}_n - \Delta.\mathcal{V}^{\prime}(v_n)=0
\label{dyndamp2}
\end{eqnarray}
where  operator $\Delta$ is defined by $\Delta. F_n=F_{n+1}+F_{n-1}-F_n$ and
$\gamma>0$ is the damping constant.
 A solution of eq.\ref{dyndamp2} is $v_n(t)=v$ where  $v$ is an arbitrary constant .
The linearized equations for  $|v_n(t)-v|$ small at  $\gamma=0$,
yield plane wave solutions $v_n(t)=v+ A \cos (qn-\omega(q) t-\alpha)$
where $\omega^2(q)= 4 \mathcal{V}^{\prime \prime}(v) \sin^2 q/2 \approx s^2(v) q^2$ for small $q$.
For avoiding more complex  situations where the front breaks into several fronts ( phase separation),
it is convenient (and physically reasonable) to assume that the square of the sound velocity $s^2(v)= \mathcal{V}^{\prime \prime}(v)$ is  a monotone decreasing function of $v$ that is $\mathcal{V}^{\prime\prime\prime}(v) <0$.

\section{Continuous Model}
Since continuous model are often used for describing fluids, we investigate first the continuous version of this model which will appear physically inconsistent but
nevertheless will reveal interesting features. Assuming  $|u_{n+1}-u_n|$ small that is $u_{n}(t)$ is a slowly varying function of $n=x$ (while the variation of $u_n(t)$ is not necessarily small),  the PDE
 \begin{eqnarray}
&\frac{\partial^2 u}{\partial t^2} &-\gamma \frac{\partial^3 u}{\partial x^2 \partial t} -\frac{\partial \mathcal{V}^{\prime}(\frac{\partial u}{\partial x})}{\partial x}=0 \qquad \mbox {or equivalently}  \label{contapp}\\
&\frac{\partial^2 v}{\partial t^2} &-\gamma \frac{\partial^3 v}{\partial x^2 \partial t} -\frac{\partial^2 \mathcal{V}^{\prime}(v)}{\partial x^2}=0   \label{contapp1}
\end{eqnarray}
describes  $u_n(t)=u(x,t)$ or $v_n(t)=v(x,t)=\frac{\partial u}{\partial x}$.
Eq.\ref{contapp1}  exhibits  exact step front solutions
$v(x,t)= v_{-\infty} + (v_{+\infty}-v_{-\infty}) \mathbf{Y} (x-ct)$ (with fast variation!)
corresponding  to $u(x,t)=v_{-\infty} x+ (v_{+\infty}-v_{-\infty})(x-ct) \mathbf{Y}(x-ct)$.
($\mathbf{Y}(x)$ is the Heavyside function ($\mathbf{Y}(x)=0$ for $x<0$
and $\mathbf{Y}(x)=1$ for $x>0$). The square of the front  velocity $c$
\begin{equation}
c^2(v_{-\infty},v_{+\infty})
= \frac{\mathcal{V}^{\prime}(v_{+\infty})-\mathcal{V}^{\prime}(v_{-\infty})}{v_{+\infty}-v_{-\infty}}
\label{frtvel}
\end{equation}
is a only function of the atomic compression  at infinities $v_{+\infty}$   and
$v_{-\infty}$. The difference $V_P=\dot{u}(-\infty,t)-\dot{u}(+\infty,t)$ is the  impact velocity
\begin{equation}
V_P= c(v_{+\infty}-v_{-\infty})
\label{impvel}
\end{equation}

Considering the energy of a finite but long part of the chain containing the front $\Phi(t)=\int_{-L}^{+L} \left( \frac{1}{2} \dot{u}^2 + \mathcal{V}(\frac{\partial u}{\partial x})\right) dx$ , we readily obtain for  any solution $u(x,t)$ of eq.\ref{contapp} with smooth second derivatives that the rate of energy variation of the system per unit time  $\dot{\Phi}= \dot{\Phi}_0$ where $\dot{\Phi}_0=\mathcal{V}^{\prime}(u(L,t)) \dot{u}(L,t) - \mathcal{V}^{\prime}(u(-L,t) \dot{u} (-L,t) $ is
the power delivered by the pressure at the edge of the system (energy conservation).
When the second derivatives of $u(x,t)$ are not smooth but involve Dirac functions, there
is generally no  energy conservation!   An explicit calculation of the energy $\Phi$
for the step front solution  (using eq. (\ref{frtvel}))  yields the dissipated power
 \begin{equation}
D= \dot{\Phi}-\dot{\Phi}_0= -c \left(\mathcal{V}(v_{+\infty}) - \mathcal{V}(v_{-\infty})
-\frac{1}{2} (v_{+\infty}-v_{-\infty})(\mathcal{V}^{\prime}(v_{+\infty})+\mathcal{V}^{\prime}(v_{-\infty})) \right)
\label{dispow}
\end{equation}
which may be written again in a simpler form as
\begin{equation}
D= \frac{c}{4} (v_{+\infty}-v_{-\infty})^3 S(v_{+\infty},v_{-\infty})
\label{disspow}
\end{equation}
where $S(x,y)= \frac{\partial R(x,y)}{\partial x} $ is the x-derivative of $R(x,y)$ defined by the equation
$\mathcal{V}(x)=\mathcal{V}(y)+(x-y)\mathcal{V}^{\prime}(y)
+\frac{1}{2} (x-y)^2 \left( \mathcal{V}^{\prime \prime}(y)+R(x,y)\right)$.
We remark that
$$R(x,y) = \frac{x-y}{3}\int_y^x \left[\int_y^{\xi_1} \left( \int_y^{\xi_2} \mathcal{V}^{\prime \prime \prime}(\xi) d \xi \right)d\xi_2 \right]d\xi_1$$  is non vanishing only when potential $\mathcal{V}$ is anharmonic. Then,  the above assumption
$\mathcal{V}^{\prime\prime\prime}(x)$ negative readily  implies $(x-y) R(x,y) <0$ and $(x-y)S(x,y)<0$. Thus, we have spontaneous energy dissipation when  $c( v_{+\infty}-v_{\infty})>0$ or  energy creation in the opposite case!

Because of this lack of energy conservation, this continuous model  is not physically acceptable for a correct description of impacts. Where this energy would go (or come from)?
This continuous equation at zero damping has another flaw we do not detail here.
There are smooth initial conditions which  spontaneously  develop a singularity  (a divergence of the derivative $\frac{\partial v}{\partial x} (x,t)$) within a finite time by the standard Rankine-Hugoniot mechanism. Actually,  the pressure fronts in lattice models without damping
 are not steplike but exhibit a puzzling behavior with an expanding  intermediate region and
 backward oscillations extending to $-\infty$ where nearest neighbor atoms  are in antiphase   \cite{HL78}. Despite the inconsistencies of this continuous model,
we prove now that  formula \ref{impvel}, \ref{frtvel} and \ref{dispow} which relates
  the impact velocity  $V_P= -\dot{u}_{+\infty}+\dot{u}_{-\infty}$, the compression $v_{+\infty}$ and $v_{-\infty}$ at infinity, the front velocity $c$ and the power dissipated by the front, are the same for
  the discrete case when \textit{damping is present but independently of this damping}.

\section{Pressure Fronts in Discrete Model: Exact Relations }

We assume in the discrete model \ref{dyndamp1} with $\gamma>0$, there are  travelling front solution
$v_n(t)=g(n-ct)$ described by a smooth steplike hull  function  $g(x)$  fulfilling
$\lim_{x\to \pm\infty} g(x) = v_{\pm\infty}$ ,
$\lim_{x\to \pm \infty} g^{\prime}(x)=0$ and
\begin{equation}
c^2 g^{\prime\prime}  + \gamma c \Delta.g^{\prime}(x)-
\Delta. \mathcal{V}^{\prime}(g(x))=0
\label{dyndamp2h}
\end{equation}
(Operator $\Delta$ is defined by $\Delta.F(x)= F(x+1)+F(x-1)-2F(x)$).

For proving eq.\ref{impvel}, we consider the length $L_{-N,+N}(t)=u_N(t)-u_{-N}(t) =\sum_{-N}^{N-1} v_n(t ) =\sum_{- N}^{N-1} g(n-ct )$ of  the chain between far sites $-N$ and $+N$.
We have  $\lim_{N\to+\infty} \dot{L}_{-N,+N}(t ) = \lim_{N\to+\infty}( \dot{u}_{+N} - \dot{u}_{-N})
=\dot{u}_{+\infty}-\dot{u}_{-\infty}=-V_P $. Otherwise, because of the existence of
the hull function $g(x)$, the rate of variation of the length
$c\left(L_{-N,+N}(t+1/c)-L_{-N,+N}(t)\right)$  between time $t$ and $t +1/c$ is
also equal to $ c(-g(N-ct ) +g(-N-1-ct ))=c(-v_N+v_{-N-1})$. Then,
 $\lim_{N\to +\infty} \left<\dot{L}_{-N,+N}(t )\right>=-c(v_{+\infty}-v_{-\infty})=-V_P$
 which yields  eq.\ref{impvel}.

 For proving eq. \ref{frtvel}, we consider the momentum $M_{-N,+N}(t ) = \sum_{-N}^{+N} \dot{u}_n(t )$
  of the chain between far sites $-N$ and $+N$.  Eq.\ref{dyndamp1} yields
   $\dot{M}_{-N,+N}= \gamma (\dot{v}_N-\dot{v}_{-N-1})+\mathcal{V}^{\prime}(v_N)-\mathcal{V}^{\prime}(v_{-N-1})$ and then  $\dot{M}=\lim_{N\to +\infty} \dot{M}_{-N,+N} = \mathcal{V}^{\prime}(v_{+\infty})-\mathcal{V}^{\prime}(v_{-\infty})$.
   Eq.\ref{dyndamp1} also yields $\ddot{u}_n=G(n-ct)$ where $G(x)= -c \gamma (g^{\prime}(x)-g^{\prime}(x-1))+\mathcal{V}^{\prime}(g(x))-\mathcal{V}^{\prime}(g(x-1))$. Time integration yields $\dot{u}_n= F(n-ct) + a_n $ where $F(x)=-G^{\prime}(x)/c$ and $a_n$ is time-constant.  Since $\dot{v}_n=-cg^{\prime}(n-ct)=\dot{u}_{n+1}-\dot{u}_n =F(n+1-ct)-F(n-ct)+a_{n+1}-a_n$, it comes that $a_{n+1}-a_n$
which is time constant, should be only a function of $n-ct$. It is thus also independent of $n$.
Then   it comes out $a_{n+1}-a_n=0$  when considering the limits $n \to \pm \infty$.
Thus, the rate of variation over
the period of time $1/c$  of the momentum $c(M_{-N,+N}(t+1/c )-M_{-N,+N}(t ) =
c \sum_{-N}^{+N} (F(n-1-ct)-F(n-ct))=c( -F(N-ct)+F(-N-1-ct))=c(-\dot{u}_{+N}+\dot{u}_{-N-1})$
becomes equal to $c V_P$ when $N\to +\infty$. It should be equal to  $\dot{M}
=\lim_{N\to +\infty} \dot{M}_{-N,+N}
= \mathcal{V}^{\prime}(v_{+\infty})-\mathcal{V}^{\prime}(v_{-\infty})$ which is time constant
when $N\to+\infty$. Combining this equality
with (\ref{impvel}) yields eq.\ref{frtvel}.

Finally, for proving eq.\ref{dispow},  we consider the energy of  the chain $\Phi_{-N,+N}=\sum_{n=-N}^{N} \frac{1}{2} \dot{u}_n^2 +\sum_{-N}^{N-1} \mathcal{V} (u_{n+1}-u_n)$ between sites $-N$ and $+N$.
We readily obtain $\dot{\Phi}_{-N,+N}= \sum_{n=-N}^{N}  \left(\ddot{u}_n- \mathcal{V}^{\prime}(u_{n+1}-u_n)+\mathcal{V}^{\prime}(u_{n}-u_{n-1})\right)\dot{u}_{n}
+\mathcal{V}^{\prime}(u_{N+1}-u_N)\dot{u}_N-\mathcal{V}^{\prime}(u_{-N}-u_{-N-1})\dot{u}_{-N}$.
We define $\dot{\Phi}_{-N,+N}^{(0)}=\mathcal{V}^{\prime}(u_{N+1}-u_N)\dot{u}_N-\mathcal{V}^{\prime}(u_{-N}-u_{-N-1})\dot{u}_{-N}$ as the power provided by the external pressure to  the finite chain.
We have for the infinite chain
$\dot{\Phi}^{(0)}=\lim_{N\to+\infty}\dot{\Phi}_{-N,+N}^{(0)} =\mathcal{V}^{\prime}(v_{+\infty})\dot{u}_{+\infty}-\mathcal{V}^{\prime}(v_{-\infty})\dot{u}_{-\infty}$.
 Then, using  eq.\ref{dyndamp1}, we readily obtain  $\dot{\Phi}_{-N,+N}-\dot{\Phi}_{-N,+N}^{(0)}
 =- \gamma \sum_{n=-N}^{N-1}   (\dot{u}_{n+1}-\dot{u}_n)^2
+ (\dot{u}_{N+1}-\dot{u}_N)\dot{u}_{N}-(\dot{u}_{-N}-\dot{u}_{-N-1})\dot{u}_{-N}$
Since $\lim_{N\to +\infty} \dot{u}_{\pm N}=0$, we obtain that the  power
 dissipated by the damping force in the infinite system
 $D=\lim_{N \to +\infty} (\dot{\Phi}_{-N,+N}-\Phi_{-N,+N}^{(0)}) <0$
is
\begin{equation}
D=- \gamma \sum_{-\infty}^{+\infty}   (\dot{u}_{n+1}-\dot{u}_n)^2
\label{diisspw}
\end{equation}
This dissipate power may be calculated differently since we assume that the solution is described by a hull function  $v_n=g(n-ct)$ implying  $\dot{u}_n=F(n-ct)$. Then, we have $\Phi_{-N,+N}(t)=\sum_{n=-N}^{N}
\frac{1}{2} F^{2}(n-ct)+\sum_{n=-N}^{N-1} \mathcal{V} (g(n-ct))$
so that we readily obtained similarly as above,
the rate of variation over  an interval of time $1/c$ of the energy
 $c(\Phi_{-N,+N}(t+1/c)-\Phi_{-N,+N}(t))=
c(\frac{1}{2}(F^{2}(-N-1-ct)-F^{2}(N-ct))
+\mathcal{V} (g(-N-1-ct))-\mathcal{V} (g(N-1-ct))$ which for $N\to +\infty$ becomes
$c( \frac{1}{2} (\dot{u}_{-\infty}^2-\dot{u}_{+\infty}^2)+\mathcal{V} (v_{-\infty})-\mathcal{V} (v_{+\infty}))
=\dot{\Phi}=\lim_{N \to +\infty}\dot{\Phi}_{-N,+N}$
 which is time constant. Consequently, we obtain for
the infinite chain,
$D= \dot{\Phi} - \dot{\Phi}^{(0)} = c( \frac{1}{2} (\dot{u}_{-\infty}-\dot{u}_{+\infty})
(\dot{u}_{-\infty}+\dot{u}_{+\infty})+\mathcal{V} (v_{-\infty})-\mathcal{V} (v_{+\infty}))
 -\mathcal{V}^{\prime}(v_{+\infty})\dot{u}_{+\infty}+\mathcal{V}^{\prime}(v_{-\infty})\dot{u}_{-\infty}$.
 Using  $V_P=\dot{u}_{-\infty}-\dot{u}_{+\infty} $ and eqs.\ref{impvel} and \ref{frtvel},
 the dissipated power $D$ defined by eq.\ref{diisspw}, is found to fulfill eq.\ref{dispow}.

As a consequence, there is no stationary travelling  front  solutions $g(x)$
in the harmonic system with damping $\gamma \neq 0$ because eq.\ref{dispow} yields $D=0$
while eq.\ref{diisspw} yields $D\neq 0$. There is also no stationary travelling  front  solutions when the system is anharmonic and $\gamma=0$ because   eq.\ref{dispow} yields $D\neq 0$
while eq.\ref{diisspw} yields $D= 0$. Otherwise, since $D$ has to be negative,  front
solutions may only exist when they propagate from larger toward the lower  pressure region
 ($c>0$ when $v_{-\infty}<v_{+\infty}$).

Assuming the existence of a hull function, $g(x)$, it is straightforward to extend the same proofs for
formula \ref{impvel}, \ref{frtvel} and \ref{dispow} to any other kind of damping forces \textit{preserving
the uniform motion} of the chain for example the Abrahams-Lorentz force  proportional to the third time derivative $\dddot{u}_n$.

\section{Numerical Calculation of Pressure Fronts}
\begin{figure}
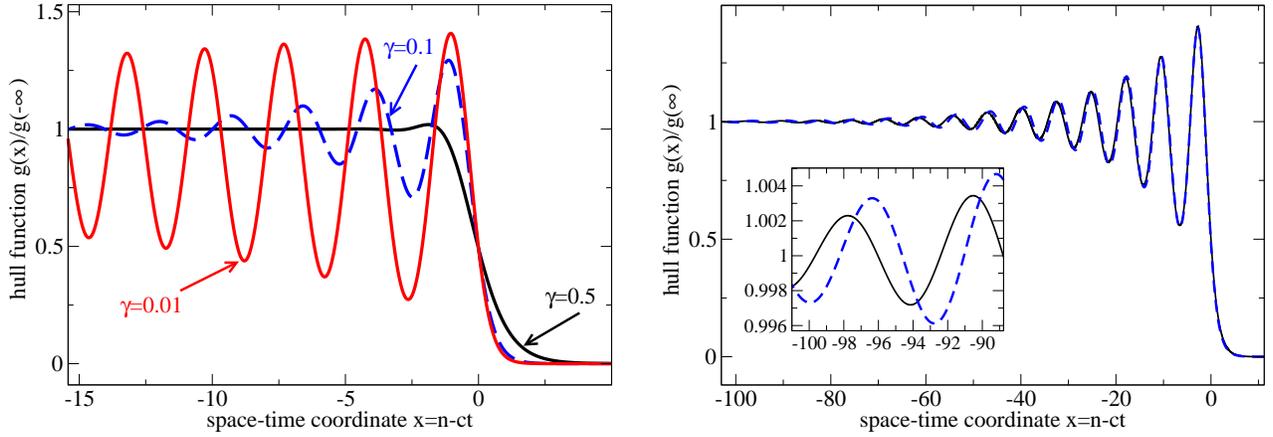

\includegraphics[width=.45\textwidth]{Fig1.eps}\hspace{0.5cm}
\includegraphics[width=.45\textwidth]{Fig2.eps}
\caption{left: Hull functions $g(x)$  for potential $\mathcal{V}(v)=\frac{1}{2} v^2- \frac{1}{6} v^3$,
 $v_{+\infty}=0$ at impact velocity  $V_P=5$ and several damping constants $\gamma$;
 right: fit (dotted line) at $\gamma=0.01$ with eq. \ref{eqfit}; insert: magnification of a part of the tail. }
\label{fig1}
\end{figure}

 The hull function $g(x)$  at nonvanishing damping  can be numerically calculated
 at computer accuracy. We start form  initial conditions where a half chain is at rest
$u_n(0)= n v_{+\infty}$ and $\dot{u}_n(0)=0$ for $n>0$ while the edge site $u_0(t)=V_P t$
is constrained to have a uniform motion at velocity $V_P>0$ (piston or impact velocity).
The front velocity $c(V_P,v_{+\infty}$  is  determined as a function of the impact velocity $V_P$ and
$v_{+\infty}$  through  eqs.\ref{frtvel}  and
\ref{impvel} which yields the implicit equation $c=\left(\mathcal{V}^{\prime}(v_{+\infty})-\mathcal{V}^{\prime}(v_{+\infty}-V_P/c)\right)/V_P$.  It is a monotone increasing function
of  $V_P$ because $\mathcal{V}^{\prime\prime}$ is assumed to be monotone
decreasing. At $V_P=0$, $c^2(0,v_{+\infty})=\mathcal{V}^{\prime\prime} (v_{+\infty})=s^2(v_{+\infty}$
becomes  the sound square velocity. The map
$$\left( \begin{array}{c }\{ v_{n+1}(1/c) \}  \\ \{\dot{v}_{n+1}(1/c)\} \end{array}\right)=\mathcal{T}
\left( \begin{array}{c }\{ v_n(0) \}  \\ \{\dot{v}_n(0)\} \end{array}\right)$$  defined by integration
of eq.\ref{dyndamp1} or \ref{dyndamp2}  over the period of time $1/c$ and a shift of the indices.
is iterated by numerically from the impact initial conditions $X$ defined above. It is found that when the damping constant $\gamma>0$ is non zero, $\mathcal{T}^p(X)$
systematically  converges for $p \to +\infty$ to a fixed point which corresponds to a solution
$v_n(t)=g(n-ct)$. Plots of an example of calculation of this hull function is shown fig.\ref{fig1}
for  several damping constants.
At strong damping $\gamma \approx > 1$,  convergence is obtained within few iterations
only and the hull function is step like. For smaller damping, the rate of
convergence slows down while the stationary  front solution develops backward oscillations which diverges at $\gamma=0$. At zero damping, there is no convergence at all.  Actually, this problem was already investigated in the literature \cite{HL78}.

At nonvanishing damping, $g(x)$  may be well approximate as a solution of a differential equation
describing an anharmonic oscillator with damping.
We apply operator $\mathbf{1}-\mathbf{Q}$ where  $\mathbf{Q}=
Q(\frac{d.}{d x})$ with  $Q(z)=1- \frac{z^2}{2(\cosh z-1)}=z^2\frac{ \sum_{n\geq 0} z^{2n}/(2(n+2))! }{\sum_{n\geq 0} z^{2n}/(2(n+1))!} =\sum_{p>0} q_n z^{2n}= \frac{z^2}{12} (1-\frac{1}{20} z^2 +....)$
 to the left member of eq.\ref{dyndamp2h}.
Since $\Delta=L(\frac{d.}{d x} )$ where $L(z)= 2 (\cosh z -1)= \frac{z^2}{1-Q(z)}$, eq.\ref{dyndamp2h} becomes $\frac{d^2.}{d x^2} \left(c^2(1- \mathbf{Q}).g +\gamma c g^{\prime} -\mathcal{V}^{\prime}(g)\right) =0$
and after two integrations
$c^2 \mathbf{Q}.g - \gamma c g^{\prime} +\mathcal{V}^{\prime}(g)-c^2 g=ax+b$
where $a$ and $b$ are two arbitrary constants. Since we search for (physical) solutions $g(x)$
which are bounded at $\pm \infty$, we must have $a=0$.  Then, defining
potential  $\mathcal{W}(g)=\mathcal{V}(g)-\frac{1}{2} c^2 g^2 -b g$, eq.\ref{dyndamp2h} takes the form
\begin{eqnarray}
c^2 \mathbf{Q}. g - \gamma c g^{\prime} +\mathcal{W}^{\prime}(g)&=&0  \qquad \mbox{or}  \label{eqhull}\\
 \frac{c^2}{12} g^{\prime \prime} - \gamma c g^{\prime} +\mathcal{W}^{\prime}(g)&\approx & 0
  \label{eqfit}
\end{eqnarray}
for the  lowest order approximation $Q(z)\approx \frac{1}{12} z^2$.
Eq.\ref{eqfit} may be viewed as the equation of a negatively damped particle with
coefficient $-\gamma c$ and mass $c^2/12$ in the effective potential $\mathcal{W}(g)$.
There are non diverging solutions for $x\to \pm \infty$ only when $\mathcal{W}(g)$ has at least two extrema and then the solution is asymptote to each of these extrema.
Actually  since $\mathcal{W}^{\prime\prime}(g)=\mathcal{V}^{\prime\prime}(g)-c^2$ is monotone decreasing, $\mathcal{W}(g)$  has at most  a maximum at $g=v_{+\infty}>v_{-\infty}$ and a minimum
at $g=v_{-\infty}$ which are determined by $b$ and $c^2$ and then eq.\ref{frtvel} is fulfilled.
Fig.\ref{fig1} (right) shows a fit of the hull function  obtained as the separatrix solution of  this equation \ref{eqfit} such that  $\lim_{x\to -\infty} g(x)=v_{-\infty}$ is the minimum of $\mathcal{W}(x)$
and $\lim_{x\to +\infty} g(x)=v_{+\infty}$ is the maximum. The error is negligible at large $\gamma$
but appears mostly as a phase shift in the tail for small $\gamma$ visible fig.\ref{fig1} right.

In summary, the most important  result of this paper is that  the energy dissipated at an impact
is independent of  the physical origin of the damping and of its value providing it preserves the translational motion of the system (as it should in physical models).
It only depends through formula (\ref{disspow}) on the anharmonic part of the potential
between  the compressions  ahead and backward the front and is proportional to the front velocity.
This result would  formally determine the emitted power of sonoluminescence if one believes
our physical interpretation. It is then straightforward  to check that the emitted power is negligible
for nearly harmonic impacts but in principle could  approach 100\% of the input power  when $\mathcal{V}$ is a hardcore potential diverging at some $v_c<v$  and for strong impacts
where $ v_{-\infty} \approx v_c$.

\bibliographystyle{ws-procs9x6}
\bibliography{ws-pro-sample}
\end{document}